# The Gluon Exchange Model for diffractive and inelastic collisions

Marek Jeżabek[1] and Andrzej Rybicki[1]

[1] Institute of Nuclear Physics, Polish Academy of Sciences,
Radzikowskiego 152, 31-342 Kraków, Poland

**Abstract**

We propose a new model for a homogeneous description of hadron-hadron and hadron-nucleus collisions, the Gluon Exchange Model, fundamentally based on color octet (gluon) exchange. In proton-proton collisions we provide an exact description of the final state proton and neutron spectrum including the proton diffractive peak. In proton-nucleus reactions we find that the projectile proton diquark cannot survive in more than about half of multiple proton-nucleon processes and consequently must be very frequently disintegrated, leading to long transfers of baryon number over rapidity space.

**1. Formulation of the model**

In this letter we introduce a new model for soft processes, mainly aimed at studies of transport of baryon number (baryon stopping) in hadron-hadron and hadron-nucleus collisions. This new Gluon Exchange Model (GEM) can be regarded as a natural generalization of the original Dual Parton Model (DPM) from which we inherit most of the formalism, see Refs. [1,2] for a detailed description. The key quantity in GEM is the number of exchanged color octets (gluons); the Fock space of states available for the participating protons/nucleons is significantly, even if naturally, extended when compared to DPM. An inherent feature of our model is that it provides a fully homogeneous description of the *entire* baryon spectrum, including fast final state protons usually known as the "diffractive peak". This we consider as very important because the "diffractive" region of the proton distribution was up to now considered as originating from completely different dynamical mechanisms, not attributable to color exchange (see, e.g., Ref. [3]). The latter applies even to our very recent work where we demonstrated the necessity of diquark disintegration in SPS proton-nucleus data [4], a subject which we will also address in the present letter.

In DPM soft interactions are described as two-step processes. At first the colliding particles exchange soft gluons and as a result, they can be described as systems composed of colored constituents. The number of these constituents depends on the number of two-body collisions of a hadron scattered on another hadron or on an atomic nucleus. In particular it is assumed that for nucleon-nucleon collisions each particle is composed of a valence diquark D and a valence quark q, and it can be described by a state |Dq> in the Fock space. The valence constituents are carrying flavors of the hadrons in the initial state. For a multiple scattering on *n* wounded nucleons in a nucleus the scattered nucleon is described by a state in the Fock space

$$| \Psi_{1,n}> = |Dq \ q_s\bar{q}_s ....q_s\bar{q}_s > \qquad \text{with } n-1 \text{ pairs} \quad q_s\bar{q}_s \ , \qquad (1)$$

corresponding to the valence diquark, the valence quark, and *n*-1 sea quark-antiquark pairs. The constituents are carrying color charges: quarks are color triplets **3**, whereas diquarks and antiquarks are color antitriplets **3**$^*$. As each of the $n$ wounded nucleons in the nucleus participates in only one two-body interaction, their state $|\Psi_{n\otimes 1}\rangle$ is described by a tensor product

$$|\Psi_{n\otimes 1}\rangle = |Dq\rangle \otimes |Dq\rangle \ldots \otimes |Dq\rangle \qquad (n \text{ times}) , \qquad (2)$$

i.e., it is composed of $n$ valence diquarks and $n$ valence quarks.

In the following discussion the $2n$ constituents of the wounded nucleons are moving to the left and $2n$ constituents of the (beam) nucleon are moving to the right in the two-body collision center-of-mass frame. DPM assumes that in the second step $2n$ highly excited string-like hadronic states (chains) are formed which fragment into hadrons in the final state. Each of the chains contains one of the left-moving constituents and one of the right-moving ones, i.e. in our case there is one D-q chain, containing the right-moving diquark and a left-moving valence quark, a q-D chain, $n-1$ chains of $q_s$-D type, and also $n-1$ chains of $\bar{q}_s$-q type. If baryon pair production is neglected (or subtracted from the studied distribution), the chains of $\bar{q}_s$-q type do not contribute to the distribution of baryons in the final state, whereas the D-q chain gives the dominant contribution to the production of right-moving baryons, and the q-D and $q_s$-D chains contribute mainly to left-moving ones. It has been shown [4] that DPM predicts too hard spectra of secondary baryons in proton-carbon collisions, and therefore has to be extended.

GEM assumes that two-body interactions are dominated by the exchange of soft color octets (gluons). For the exchange of $n$ gluons the right-movers are in a color representation

$$\mathbf{R} \subset 8 \otimes 8 \ldots \otimes 8 \qquad (n \text{ times}) \qquad (3)$$

and the left-movers are in the complex conjugate representation $\mathbf{R}^*$. It is evident that the states $|\Psi_{1,n}\rangle$ and $|\Psi_{n\otimes 1}\rangle$ correspond to a special case of Eq. (3) for $n$ wounded nucleons in the nucleus.

For the case of one gluon exchange in GEM, the state of the right moving constituents is a statistical mixture of two states:

$$|\Psi_{1,1}\rangle = |Dq\rangle$$

with probability $(1-r)$, and

$$|X_{1,1}\rangle = |[q_1,q_2,q_3]\ q_s\bar{q}_s\rangle$$

with probability $r$, where $[q_1,q_2,q_3]$ denotes three valence quarks in a color singlet (antisymmetric) state. Let us note that the states $|X_{1,1}\rangle$ and $|\Psi_{1,2}\rangle = |Dq_3\ q_s\bar{q}_s\rangle$ are orthogonal because $Dq_3$ is a system composed of the three valence quarks in the color octet state (the diquark D is a system of two valence quarks in an antisymmetric color state, $D = [q_1,q_2]$ ).

For the state $|X_{1,1}\rangle$ the color structure of the valence constituents is not changed in the color exchange process, because the corresponding gluon is attached to the pair $q_s\bar{q}_s$. This means that in the final state a fast baryon will be produced, having the same flavor as the incoming projectile baryon and separated from other secondary particles by a gap in rapidity. Such events are generally labeled as "inelastic diffraction". It is very natural to consider a more general family of states

$$|X_{1,n}\rangle = |\ [q_1,q_2,q_3]\ q_s\bar{q}_s \ldots q_s\bar{q}_s\rangle \qquad (\text{with } n \text{ sea pairs}) \qquad (4)$$

which will describe inelastic diffraction of a nucleon for the exchange of $n$ gluons. However, such contributions are suppressed by a factor $r^n$ and therefore will be neglected in the present paper.

Let us consider now the case of two exchanged gluons and two wounded nucleons in the nucleus. In the framework of GEM a new possibility arises: three valence quarks $|q_1q_2q_3>$ in the color decuplet state moving to the right, and a color antidecuplet system $|D_1D_2[q_1,q_2]>$ moving to the left, composed of two diquarks ($D_1$ and $D_2$) and two valence quarks ($q_1$ and $q_2$) in the color antitriplet state. These two valence quarks in the nucleus can be considered as an effective diquark. As a result three color singlet chains are formed, all of them of the q-D type. However, another possibility exists that four chains are formed and the right-movers are described by the next state in the Fock space $| q_1q_2[q_3,q_s] \bar{q}_s >$, where the effective diquark is composed of $q_3$ and $q_s$. In both cases the right-moving valence diquark will be disintegrated and its components (i.e., two valence quarks) will find themselves in different chains. As a result the spectrum of right moving baryons will be softer than for the DPM-like contributions. For a larger number $n$ of exchanged gluons, one can built sequences of corresponding states in a way analogous to DPM. In particular, for the right-movers one $q_s\bar{q}_s$ pair of constituents is added for any additional gluon.

We take the following constituent distribution for our GEM model:

$$\rho_m(x_{q_1}, x_{q_2}, x_{q_3}, x_1, ..., x_{2m}) =$$
$$C_m(x_{q_1} + x_{q_2})^{1/2} x_{q_3}^{-1/2} \prod_{i=1}^{2m}(x_i^2 + 4\mu^2/s)^{-1/2} \cdot \delta\left(1 - x_{q_1} - x_{q_2} - x_{q_3} - \sum_{i=1}^{2m} x_i\right), \quad (5)$$

where $C_m$ is a normalization factor, $x_{q_1}$, $x_{q_2}$, $x_{q_3}$, $x_1$, ..., $x_{2m}$ are the momentum fractions carried (in the most general case) by the three valence quarks and the $2m$ sea quarks and antiquarks, $\mu$ is the transverse mass of the sea quark, and $s$ is the collision c.m.s. energy squared. We note that the above formula (5) is a straight-forward generalization of Eq. (2.1a) from Ref. [2] and, with $x_d = x_{q_1} + x_{q_2}$, it integrates to the latter equation for any diquark-preserving sequence discussed in Secs. 3-4 below. Calculations performed on the basis of the explicit form of Eq. (5) will be presented in a more detailed paper [5].

In the situation of two incoming protons, or one proton and one nucleus, we consider the following sequences for color exchange:

* <u>For one gluon exchange</u> (pp collision, or collision of a proton with one nucleon in the nucleus):

(1) *both protons are in their basic Fock state $| \Psi_{1,1}>$, Fig. 1a;* after gluon exchange the valence Dq is in a color octet state; thus, two new chains (D-q and q-D) are formed; this is exactly the phenomenology of the Dual Parton Model which can be treated by the same formalism [1,2].

(2) *one proton is in the state $| \Psi_{1,1}>$, but the other proton is in its next Fock state $| X_{1,1} >$, Fig. 1b;* gluon exchange occurs between the valence Dq of one proton and the $q_s\bar{q}_s$ pair of the other proton; thus, the valence (uud) structure of the latter proton remains in a color singlet state, and produces one fast proton in the "diffractive peak"; the $q_s\bar{q}_s$ pair is in a color octet state, which results in the formation of two chains ($q_s$-D, $\bar{q}_s$-q).

(3) *both protons are in their next Fock state;* this sequence is suppressed by a factor of $r^2$ so we assume that it can be neglected in the present, introductory analysis (this subject will be further discussed in Sec. 3).

\* For more-than-one gluon exchange, with *n* gluons (p-nucleus collision) :

(i) *the "pure" DPM case,* where the proton is in its lowest possible Fock state $|\Psi_{1,n}>$ , and the wounded nucleons are in the basic state $|\Psi_{n\otimes 1}>$, is shown in Fig. 1c for the simplest case of *n*=2; this configuration where *n* gluons are exchanged between the valence Dq and ($n-1$) $q_s\bar{q}_s$ pairs from the projectile proton and the *n* valence Dq's from the wounded target nucleons is again fully treatable with the DPM formalism [1,2] as it was the case for the sequence (1) above.

(ii) *the "single diffraction in nucleus" case, Fig 1d,* same as above but with one nucleon from the nucleus in its next Fock state $|X_{1,1}>$ ; this leads the presence of one "diffractive" nucleon on the nucleus side.

(iii) *diquark disintegration*, where the multiple gluon exchanges bring the three valence quarks of the projectile into a *color decuplet* state, and therefore discern the full quark structure of the proton ( |qqq> + higher components of Fock space ). Such configurations, *not possible* with single gluon exchange (*n*=1), will result in the creation of color singlets made of three quarks. Examples of such diagrams are drawn in Figs. 1e and 1f. The above singlets will contribute to the final state baryon distribution in the proton hemisphere, even if this contribution will be mixed with that coming from fragmentation of chains of the q-D or $q_s$-D type (the former are also apparent in Figs. 1e, 1f).

(iv) *uud valence constituents untouched*, which is direct generalization of the sequence (2) above to *n*>1, with color octets attached only to sea quarks from the projectile as drawn in Fig. 1g. This interesting but quite strongly suppressed possibility which results in the production of forward "diffractive-like" protons in pA collisions with 3 or more participants (2 or more participating target nucleons) will be addressed in a more detailed paper [5].

As GEM keeps the two-step structure of the DPM, a comment about the 2$^{nd}$ (fragmentation) step should be made for completeness. In the present paper the latter is realized in a way basically identical to that used for the DPM in Ref. [2], with singlet fragmentation functions into protons and neutrons being (by necessity) adjusted to experimental pp data and with no extra parameters brought to the model for pA collisions.

## 2. The experimental data

Our calculations made with GEM will be compared to experimental data on pp and pC collisions obtained by the NA49 detector [6,7]. These precise data cover nearly the entire projectile hemisphere (up to $x_F$=0.95 and with no lower $p_T$-cutoff) and include both protons and neutrons. Consequently, a strong advantage of our study is the possibility to use baryon number conservation as an additional constraint in the verification of our model. We note that the integration of net p and net n spectra per pp hemisphere, supplemented by strange baryon multiplicities provided in Ref. [8], gives us unity within an accuracy better than 5%, fully compatible with experimental uncertainties [6].

In Fig. 2 we present the comparison of $x_F$-distributions of protons in pp reactions [6], minimum bias pC collisions [7], and in proton-carbon events where the projectile proton collides with more than one nucleon from the carbon target. The latter distribution is obtained assuming that the proton spectrum in single proton-nucleon collisions is similar to that in pp reactions, as

$$\frac{dn}{dx_F}(pC_{\text{multiple collisions}} \to pX) =$$

$$\frac{1}{1-P(1)}\left(\frac{dn}{dx_F}(pC \to pX) - P(1)\cdot\frac{dn}{dx_F}(pp \to pX)\right), \quad (6)$$

where P(1) is the single-collision Glauber probability taken from Ref. [9]. The resulting proton spectrum in "multiple collision" pC events[1] displays the natural increase in steepness (stronger baryon stopping) with respect to pp and minimum bias pC collisions. We also note the vanishing of the proton "diffractive peak" at high $x_F$ which confirms our expectation of a strong suppression of the sequence (iv) above which is, in our model, responsible for the emission of such protons through the preservation of the uud singlet.

### 3. A homogeneous description for "diffractive" and "inelastic" pp collisions

Fig. 3 presents the description of net proton and net neutron spectra in pp reactions which we obtained in the framework of our GEM model. In order to eliminate the contribution from baryon-antibaryon pair production from NA49 data [6] we took net proton distributions as p-$\bar{\text{p}}$ and net neutron distributions as n-$\bar{\text{n}}$, and took $\bar{\text{n}} \approx \bar{\text{p}}$ for simplicity. We derived the neutron rapidity spectrum from its published $dn/dx_F$ distribution assuming the shape of the neutron $p_T$ spectrum at a given $x_F$ to be similar to that of the proton $p_T$ spectrum.

In the present GEM simulation, we inherit the main ingredients of the original DPM formalism described in Ref. [2] including diquark/quark masses, etc., but take the sea quark mass as $\mu = 60$ MeV/$c^2$. The details of the q-D/D-q fragmentation functions into protons and neutrons (constructed following identical premises as in Ref. [2]) are adjusted to the data, and the relative probability $r$ for color exchange to connect to a higher Fock state is set to about 8%. With this in mind GEM achieves an essentially precise description of both proton and neutron distributions, including the "diffractive" proton peak at high rapidity. This we find remarkable as the latter peak is generally believed not to be caused by color exchange processes, see e.g. Ref. [3]. The present description is better than that which we obtained [4] using the pure DPM as the latter does not contain the higher Fock state diagram (2) from Fig. 1b, and even more so if compared to an earlier DPM analysis by one of us [2]. We conclude that the inclusion of a properly complete ensemble of Fock states, not present in DPM but present in GEM, is necessary for a fully successful, homogeneous description of the transport of baryon number in pp collisions. This is an issue we will further discuss for pA reactions in Sec. 4.

Specific detailed aspects of this analysis should now be commented upon. As it is evident from Eq. (5) above, the mass $\mu$ represents the cut-off in the longitudinal momentum distribution of sea quarks and antiquarks. As such it is strongly constrained by the experimental measurement of the width of the "diffractive" peak which corresponds, in the framework of GEM, to that of the distribution of the preserved (uud) valence singlet. We find the corresponding allowed range of $\mu$ not to exceed 40-100 MeV/$c^2$, with $\mu = 60$ MeV/$c^2$ giving in the best agreement with the experimental data. On the other hand, the value of the probability $r$ is dictated by the height rather than the width of the experimental diffractive peak. Our present estimate of the accuracy of $r$ is about 8±1%. Consequently we state that the contribution of $r^2 \approx 0.6\%$ events where *both* protons are in the next Fock state, which in GEM will result in a simple scaling of the (uud) peak in Fig. 3, can be neglected account taken of the present accuracy of our study. It should be stressed, however, that the relative probability $r^2$ of "double inelastic diffraction" events characterized by two fast protons separated by rapidity gaps from a central production is an evident, firm prediction of our model.

---

[1] We take the term 'multiple collisions' from Ref. [9] where a similar operation was performed for distributions of charged pions. We note the general similarity of our reasoning to that made in the work by Busza and Goldhaber [10].

**4. pA reactions**

For the collision of the projectile proton with $n>1$ nucleons, the original DPM model [2] predicted an enhanced "stopping" for the projectile diquark (and leading baryon) distribution by the mechanism connected to creation of sea quark-antiquark pairs which in this paper corresponds to the diagram in Fig. 1c. For the DPM, we demonstrated the failure of this scenario in our recent work, limited to total non-strange baryon spectra [4]. Presently we study this problem more accurately with GEM, and consider separately protons and neutrons.

Fig. 4 shows the comparison between net proton and net neutron distributions in pC collisions where the projectile proton collides with more than one nucleon from the C target, and our GEM model calculation. The former are obtained directly from NA49 data [6,7] in the same way as we did it in Sec. 2, using the Glauber simulation [9] and an equation directly analogous to (6) but taken in rapidity. The GEM calculation is limited to the diquark-preserving scenario which is taken as follows. From the three diquark-preserving diagrams in Fig. 1 (c, d, and g), the latter (g), with color octets attached only to sea quarks from the projectile is strongly suppressed (by a factor $r^n$, with $r = 8\%$ as said in Sec. 3). Consequently we disregard it here. For the two non-negligible diagrams (c, d), it can be readily shown that in the considered range of rapidity, their sum can be very well approximated by the pure diagram (c) with the normalization of the q-D type strings, connected to carbon diquarks, scaled down by a factor of $(1-r) = 0.92$. This is because the spectrum of the "diffractive" nucleon in Fig. 1d does not extend into the range of rapidity considered in Fig. 4.

It is evident from the figure that the scenario of preserving the diquark, and softening of the projectile $D$-$q$ chain underpredicts the nuclear stopping power[2] in processes where the proton collides with more than one nucleon from the carbon nucleus. In this respect the conclusions from the pure DPM and GEM coincide. However, the extension of the analysis to protons and neutrons taken separately allows for a discussion more detailed than in Ref. [4]. The precision of the experimental data allows us to establish a tentative upper limit for the contribution of this diquark-preserving scenario to the total net baryon spectra. This is illustrated by means of the dashed lines in the figure, which shows our model simulation scaled by 0.48. We note that the neutron simulation saturates the experimental data at forward rapidity and therefore conclude that the diquark-preserving diagrams (c) and (d) can account for about a half of multiple proton-nucleon processes in pC reactions. This upper limit is slightly more restrictive than the ~0.6 which we obtained with the pure DPM [4]. Interestingly, unlike for neutrons, the scaled GEM proton simulation somewhat underestimates the spectrum of protons at forward rapidity. With due caution induced by systematic errors of the NA49 neutron measurement, we tend to attribute the apparent small surplus of protons to the strongly reduced uud singlet component from the projectile proton, induced by the diagram (g) in Fig. 1.

The above is, in our view, the most accurate up to now delimitation of the contribution of diquark-preserving mechanisms to the leading baryon distribution in multiple proton-nucleon collisions. Apart from the more general formulation of GEM which results in a more complete description of baryon spectra in the pp reaction than the relatively over-restricted DPM, and consequently in a more robust starting point for this analysis, an important merit is to be attributed to the experimental data [6,7] which provide nearly hermetic information on protons and neutrons in the projectile hemisphere. What comes out from this study is that already for the pC reaction, the projectile diquark must be disintegrated in about half of collisions of the proton with two or more nucleons. As evident from Fig. 4, the resulting contribution to the baryon rapidity spectrum will be significantly softer than for the diquark-preserving case. This is evident, on the qualitative level, from diagrams (e) and (f) in Fig. 1 because both of them correspond (apart from target diquark strings of the q-D type) to color singlets which are softer than these originating from preserved projectile diquarks (Fig. 1c, d). We will present a fully quantitative study of these diagrams in a more detailed paper [5].

---

2   This term is taken from Ref. [10].

## 5. Summary

We presented a new model for soft processes in hadron-hadron and hadron-nucleus collisions, the Gluon Exchange Model. While formally our model can be regarded as a generalization of the Dual Parton Model by Capella and Tran Thanh Van, it is fundamentally based on the number of exchanged color octets (gluons) and significantly extends the Fock space of states available for the participating protons and nucleons.

In proton-proton collisions, GEM provides a complete description of the final state proton and neutron distribution in the projectile hemisphere. What is remarkable is that unlike the original DPM, see e.g. Ref. [2], GEM successfully describes the proton "diffractive peak" at high $x_F$ as a specific case of color octet exchange (Fig. 1b).

In proton-nucleus collisions where the proton interacts with more than one nucleon, our study made with GEM shows that that the projectile proton diquark cannot survive in more than about half of such processes. Consequently, the diquark must be frequently disintegrated, leading to new diagrams with long transfers of baryon number over rapidity space.

## Acknowledgments

This work was supported by the National Science Centre, Poland (grant no. 2014/14/E/ST2/00018).

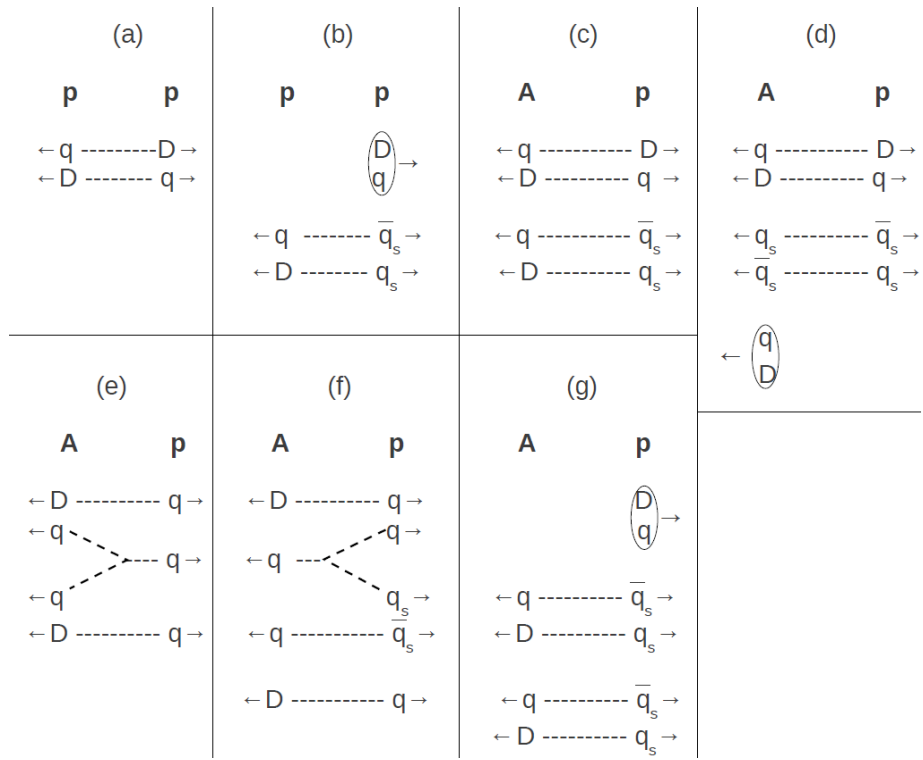

Fig. 1. The basic configurations for color octet (gluon) exchange considered in our model for one and more-than-one gluon exchange as described in the text. The dashed lines connect the newly created color singlets (chains), while the ellipse marks the final state proton resulting from the preservation of the uud valence singlet. In some of the diagrams, the different combinations for color singlet formation must evidently be taken into account.

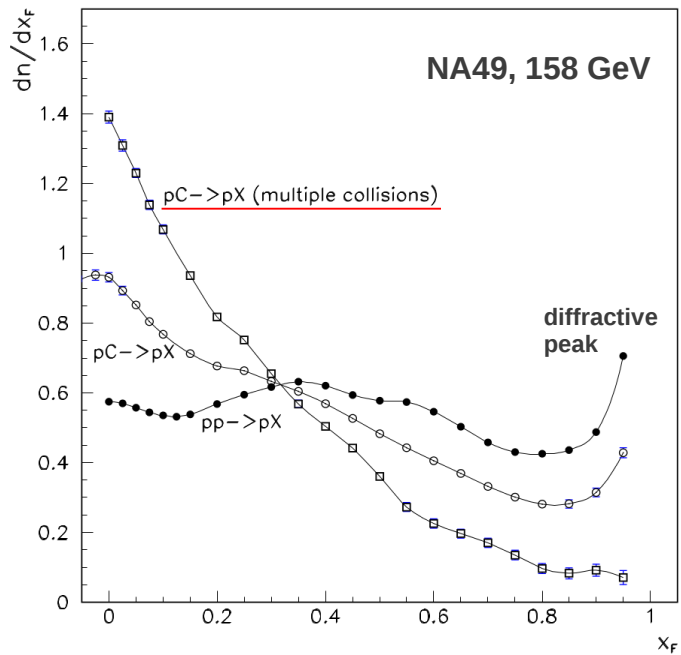

Fig. 2. Feynman-x distributions of protons in pp and minimum bias pC collisions at $\sqrt{s_{NN}}$=17.3 GeV obtained from the NA49 experiment [6,7], and in pC collisions in which the projectile proton collides with multiple target nucleons, obtained using Eq. (6) as described in the text.

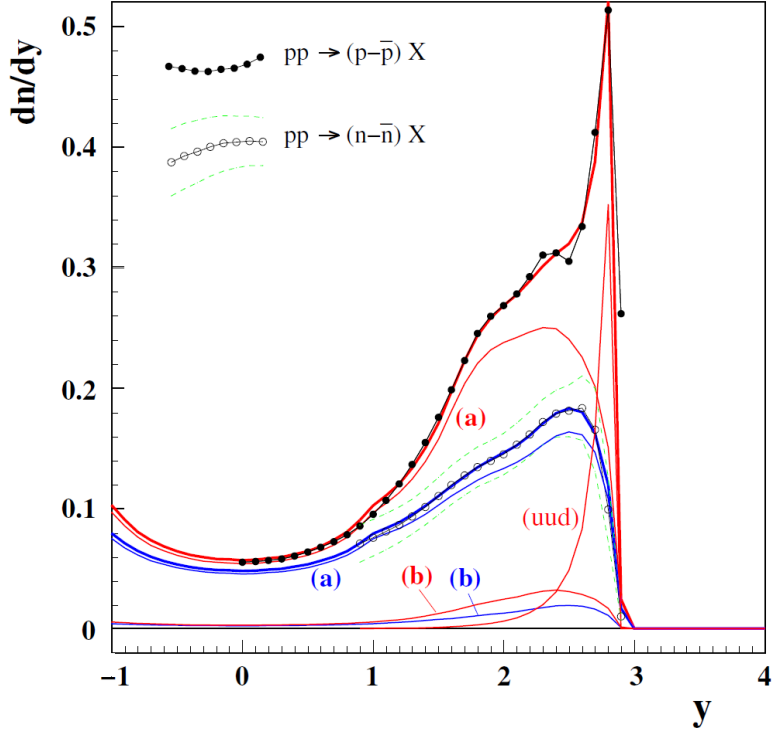

Fig. 3. Rapidity distributions of net protons and net neutrons in pp collisions at $\sqrt{s_{NN}}$=17.3 GeV obtained from the NA49 experiment [6] (data points), put together with the result of our GEM calculation (top solid red and blue lines for protons and neutrons, respectively). The contributions of the D-q (q-D) chains from the diagram (a) in Fig. 1 and of the D-$q_s$ ($q_s$-D) chains from the diagram (b) in Fig. 1 are marked as (a) and (b), respectively. The contribution of the uud singlet from Fig. 1 (b) is also indicated in the plot. The two green dashed lines reflect the systematic error of the NA49 neutron data which is the main source of uncertainty in the paper [6].

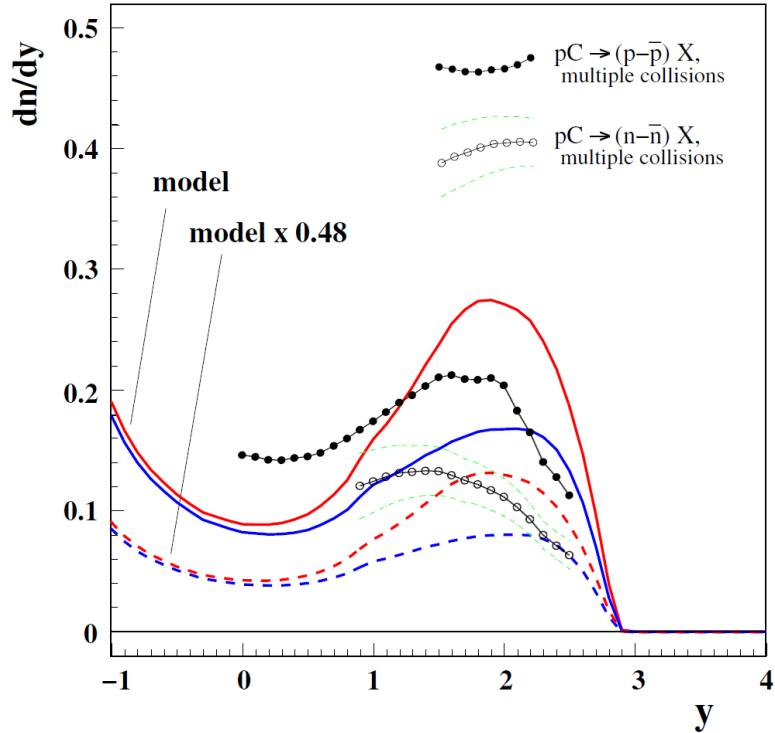

Fig. 9. (a) Rapidity distribution of net protons and net neutrons in pC reactions in which the projectile proton undergoes more than one collision with carbon target nucleons, obtained from the NA49 experiment [6,7], and compared to our GEM calculation performed for the diquark-preserving scenario as described in the text. The calculation for protons is drawn in red and for neutrons is drawn in blue. The dashed lines show the result of the same calculation scaled by 0.48. The two green dashed lines reflect the systematic error of the NA49 neutron data [6,7].